\documentclass{osa-article}

\journal{oe}

\articletype{Research Article}
\usepackage{graphicx,xcolor}
\usepackage{blindtext}
\usepackage[utf8]{inputenc}
\usepackage{multirow}

\begin{document}

\title{Multipolar scattering analysis of hybrid metal-dielectric nanostructures}

\author{Debdatta Ray,\authormark{1} Andrei Kiselev,\authormark{1} and Olivier J.F.\ Martin\authormark{*}}

\address{\authormark{1}Both authors contributed equally to this work\\
Nanophotonics and Metrology Laboratory, Swiss Federal Institute of Technology Lausanne (EPFL), CH--1015 Lausanne, Switzerland}

\email{\authormark{*}olivier.martin@epfl.ch} 



\begin{abstract}
We perform a systematic study showing the evolution of the multipoles along with the spectra for a hybrid metal-dielectric nanoantenna, a Si cylinder and an Ag disk stacked one on top of another, as its dimensions are varied one by one. We broaden our analysis to demonstrate the "magnetic light" at energies above 1 eV by varying the height of the Ag on the Si cylinder and below 1 eV by introducing insulating spacing between them. We also explore the appearance of the anapole state along with some exceptionally narrow spectral features by varying the radius of the Ag disk. 
\end{abstract}

\section{Introduction}

Metallic nanostructures can produce a strong field enhancement by the resonant oscillation of free electrons and have thus been a popular research topic for several decades \cite{Maier_book}. There are however two main limitations to those resonances: the intrinsic losses associated with metals, which limit the quality factor that can be achieved \cite{Kottmann_2000c}; and their essentially electric dipolar character, which restricts the degrees of freedom for engineering arbitrary electromagnetic responses \cite{Ekinci_2008}. These issues can be solved by replacing metals with low loss dielectrics, which have recently emerged as a vivid research topic \cite{Yang_2014,Kuznetsov_2016,Yavas_2017,Tzarouchis_2017a, Feng_2017}. Dielectric nanostructures exhibit both electric and magnetic resonances, which enable several interesting phenomena like unidirectional scattering \cite{Kiselev_2020,Achouri_2020} Fano resonances \cite{Limonov_2017}, or even anapole states that do not radiate in the far field and are formed by the destructive interference between electric and toroidal dipoles \cite{Miroshnichenko_2015}. Unfortunately, despite all their advantages, the use of dielectrics reduces the field enhancement in the vicinity of the structure, which is detrimental for applications like fluorescence \cite{Darvill_2013} and sensing \cite{Han_2019b}. In order to circumvent the drawbacks of both metals and dielectrics, hybrid metal-dielectric nanostructures have recently come up as a promising option. These nanostructures are expected to provide the best of both worlds by combining low losses from dielectric and high field enhancement from metal. 

The significant progress made in various facets of hybrid nanophotonics has been efficiently collated in a review by Lepeshov et al. \cite{Lepeshov_2019}, which illustrates experimentally achievable hybrid geometries based on different working principles like the whispering gallery modes in Si-Al\textsubscript{2}O\textsubscript{3}-Ag dimers leading to directional radiation \cite{Wang_2013} and the coupling of a hybrid mode at the Si-Ag interface of a Si nanoparticle on an Ag film to the propagating surface plasmon-polariton \cite{Yang_2017b}. Unconventional and newer hybrid core-shell geometries like an Ag rod covered by a semiconductor layer for broadband absorption enhancement \cite{Mann_2013} and a gold rod covered by a Cu\textsubscript{2}O shell for generation of third harmonic \cite{Elli_2015}, to name a few, has also been mentioned. The hybrid nanoantenna, due to the presence of dielectrics, boasts of some unique features like the pure magnetic dipole resonance or the anapole states, which will also be seen in the later sections of this work. Predominantly Magnetic Fano resonances have also been observed in hybrid oligomers formed from a combination of Si cones and Au particles, which can be tuned by laser melting of the Au particles \cite{Lepeshov_2016, Lepeshov_2017}. Additionally there are several other recent works on hybrid nanoantenna that also need to be acknowledged such as the ones that discuss the efficient generation of the second \cite{Gili_2018, Hu_2020b}, and third harmonics \cite{Shibanuma_2017}; nonlinear optical properties \cite{Hentschel_2016}, subtractive color filtering \cite{Yue_2016, Nagasaki_2018}, sensing \cite{Debdatta_2020}, unidirectional radiation\cite{Ho_2018} and asymmetric absorption \cite{Yang_2020a}.  

Upon careful considerations of these articles it can be noted that core-shell or stacked cylindrical geometries are the most popular. While the core-shell geometry has analytical solutions, the stacked cylindrical geometry still lacks a consistent theoretical tool for the analysis. One possible way to fulfill this gap is to perform a comprehensive systematic parametric study on how the geometrical parameters of the system influence the scattering characteristics of these structures, which represents the focus of this paper.  

To the author's knowledge, only a few publications have unveiled some of these dependencies. For example, the scattering characteristics of dielectric core coated with metal shell have been implemented experimentally and studied analytically for spherical and numerically for co-axial cylindrical configurations \cite{Feng_2017,Li_2020d,Xu_2019a,Liu_2012n}. A hybrid geometry with metal nanorods on top of Si cylinders was studied, noting the influence of a spacer between a metal rod and a dielectric cylinder to enhance directional radiation \cite{Rusak_2014, Guo_2016}. The effects of thickness and permittivity of the dielectric spacer and the metal nanodisk in a metal-dielectric-metal sandwiched geometry on the directional radiation and manipulation of multipoles were also investigated \cite{Zhang_2019e, Qin_2019a}. 

In principle, the coupled electric and magnetic dipole theory that works for spherical objects \cite{Sun_2020} could be applied to predict scattering from cylindrical geometries. Unfortunately, from our experience, this approach does not give sufficiently accurate results and we have to resort to numerical methods to explain how various features of these hybrid geometries appear and disappear as we perform a systematic study of one specified hybrid geometry by varying its dimensions. For our work we chose a metal-dielectric nanostructure made of Si cylinder and Ag disk placed on top of one another. We start by varying the geometrical parameters like height and radius of a Si cylinder, followed by Ag disk and finally that of a hybrid structure. Since the parameters considered here are experimentally meaningful, this systematic analysis showing how each geometrical parameter affects the spectral response is also very beneficial from an engineering perspective for designing hybrid geometry. This analysis and the underlying physics can also be extended to several other similar hybrid geometries like core-shell particle \cite{Savalev_2017, Liu_2012o, Wang_2015} and cylinder-disk \cite{Yue_2016, Nagasaki_2018} and cylinder-nanorod geometries \cite{Rusak_2014, Guo_2016}.

We perform the numerical simulations with the surface integral method, which has been proven a very accurate frequency domain technique \cite{Gallinet_2015,Kern_2009}. To avoid the singularities of the Green's tensor close to the scatterer's surface, we use the singularity subtraction technique \cite{Raziman_2015, Kern_2010}. Refs. \cite{Kern_2009} and \cite{Raziman_2015} discuss in detail the performance of this technique and the accuracy of the algorithm in computing the response from various nanostructures. Numerical simulations are challenging and choosing a method that is best suited to the needs of the analysis with minimal trade offs is always an art. In this work we have presented all the results of the parametric study using a well-established SIE method. The use of a single method throughout the paper helps in building a coherent understanding as the geometry progresses from a single Si cylinder to a more complicated Ag disk-Si cylinder hybrid nanoantenna. The Cartesian multipoles decomposition is performed with the vector spherical harmonic basis \cite{Muhlig_2011}, which is available online \cite{Kiselev_VSH, Kiselev_2019, Yan_2017}. The resultant multipole for any structure strongly depends on the choice of origin of the coordinate system \cite{Burko_2007}. We make the choice of origin at the geometrical center of the mesh to provide a coherent analysis. Aiming at clarifying how different geometrical parameters affect the overall spectrum, we vary the size of each component and monitor the evolution of the multipolar response traces. The system is illuminated with a planewave propagating along the symmetry axis $\textbf{z}$, leading to a polarization insensitive scattering response. Such a geometry can be fabricated with a high level of control \cite{Yang_2020a,Debdatta_2020}. We have chosen Si as the dielectric material since it has high permittivity and modest losses in the visible and has been widely used experimentally \cite{Kuznetsov_2016, Decker_2015, Staude_2017} and Ag as the metal for its low losses \cite{West_2010,Dastmalchi_2016a,Khurgin_2017}. For our simulations we use available data for Si from Ref.\cite{Pierce_1972}. To put an effort to be consistent with our recent experimental research on hybrid antenna for sensing, we use our experimentally measured data for Ag \cite{Ag_data_Debdatta} that, in comparison to the data from Johnson and Christy \cite{Johnson_and_Christy_1972}, show almost the same real part and slightly lower imaginary part of the dielectric permittivity of silver in the region 1100-1700 nm. 

\section{Results and discussion}

The results are divided into several sections to explore in details the influence of the different geometrical parameters on the spectral features supported by the system. We first explore the scattering cross-section (SCS) for a bare Si cylinder, then add a metal disk of varying thickness and diameter. Finally, we investigate the coupling between the dielectric and metallic parts as a function of their separation.

\subsection{Response of the dielectric cylinder}

To understand the advantages of a hybrid nanoantenna and the additional control it offers, we begin by analyzing the SCS of an isolated Si cylinder illuminated with a planewave. Although this has been reported by others \cite{Staude_2013,Evlyukhin_2011,Miroshnichenko_2015, Groep_2013,Haar_2016}, it will help us form a coherent understanding as we proceed to the subsequent sections, where we modify the geometry gradually by adding Ag to the Si cylinder.

\begin{figure}[h!]
\centering
  \includegraphics[width=1\textwidth]{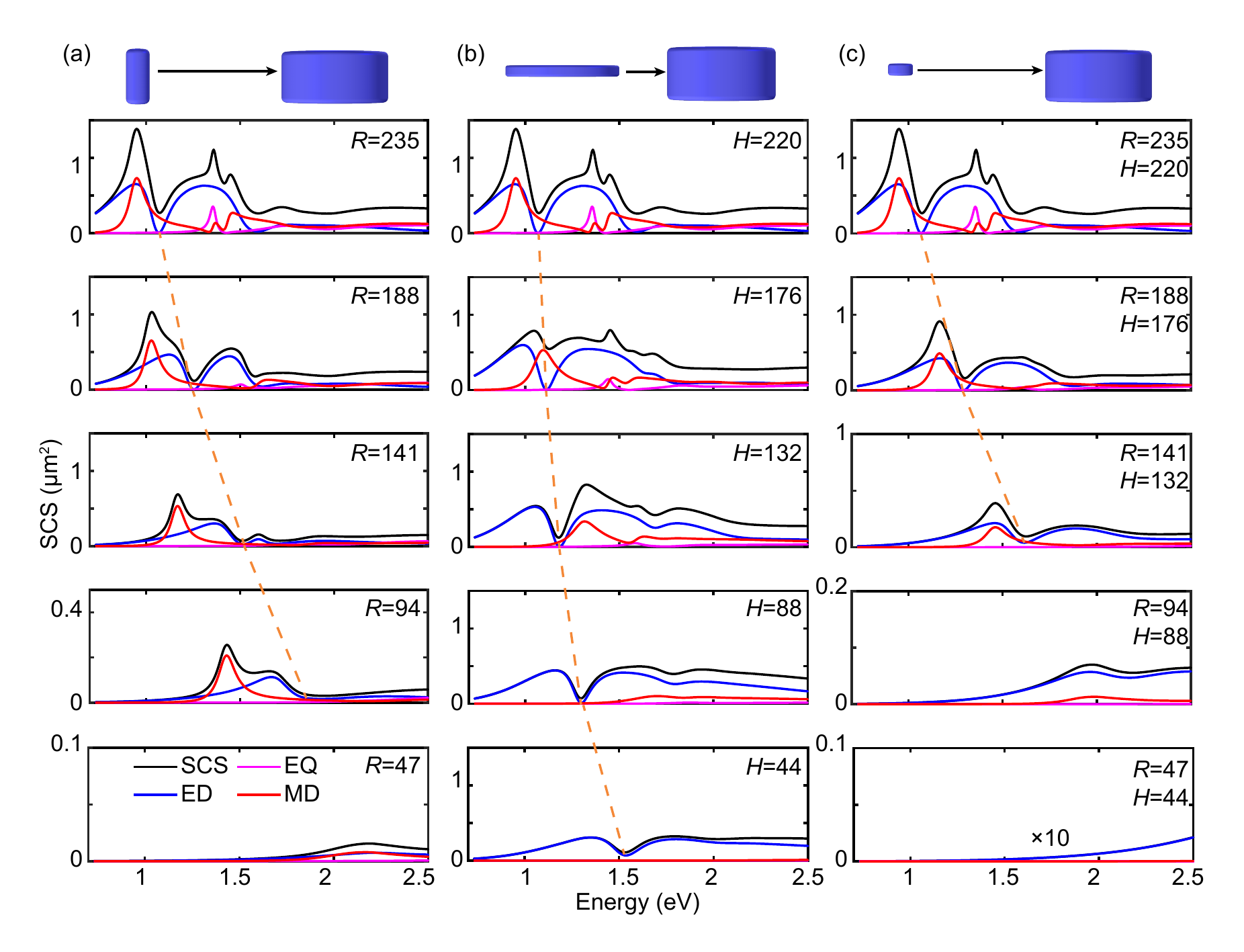}
  \caption{Scattering cross section (SCS) and its decomposition into Cartesian multipoles: electric dipole (ED), magnetic dipole (MD) and electric quadrupole (EQ). (a) Constant height $H=220$ nm and varying radius $R$, (b) constant radius $R=235$ nm and varying height $H$, (c) varying height $H$ and radius $R$. All dimensions are in nm and the orange dashed line traces the position of the anapole state.}
\label{Si_R_and_H_vary}
\end{figure}

We first vary the radius $R$ of the Si cylinder from 235 nm until 47 nm in steps of 47 nm, as shown in Fig.~\ref{Si_R_and_H_vary}(a); for the time being, we retain the height $H$ of the Si cylinder as 220 nm. Already in this simple system, we observe a rather rich spectrum with many different features, especially for the larger cylinders. The entire SCS can be decomposed into an electric dipole (ED), a magnetic dipole (MD) and an electric quadrupole (EQ) Cartesian multipoles. It can be observed that with the decrease in radius, both the ED and MD blueshift. Interestingly, the ED is shifting more than the MD, such that the ED leaves the MD far away behind, on the left side of the spectrum, as the radius decreases, as seen for $R=94$ nm in Fig.~\ref{Si_R_and_H_vary}(a). This faster shift of the ED is actually not surprising and has been observed previously for the case of a dielectric sphere (see e.g. Fig. 1(c) in Ref. \cite{Lepeshov_2019}), and is explained by analyzing the electric and magnetic Mie coefficients $a_1$ and $b_1$ in the framework of the Pad\'e expansion\cite{Tzarouchis_2017a}. The magnitudes of both ED peaks decrease as the cylinder shrinks, while the MD resonance retains its magnitude.

Let us now vary the height $H$ of the Si cylinder from 220 nm to 44 nm in steps of 44 nm, Fig.~\ref{Si_R_and_H_vary}(b). Again, the ED and MD resonance positions blueshift with decreasing sizes. Opposite to what was observed for a change of radius, now the ED magnitudes remain rather constant, while the MD magnitude drastically decreases as we diminish the height. Also, we notice that the position of the MD peak blueshifts much faster than the ED ones, leading to significant "overtaking" of the MD resonance over the ED, as seen for $H =132$ nm in Fig.~\ref{Si_R_and_H_vary}(b). Actually, this observation is important for engineering directional scattering \cite{Alaee_2015,Liu_2018,Kiselev_2020, Achouri_2020}, or producing a far-field response constituting of only one particular mode, e.g.\ the MD for "magnetic light" \cite{Kuznetsov_2012,Li_2020d}. The ED and MD resonances of dielectric structures can have strong spectral overlap between them; a difference in their shift rate enables their efficient spectral separation. For example, by varying the Si height, we can control the ED and MD responses and achieve a good spectral separation for $H=176$ nm, Fig.~\ref{Si_R_and_H_vary}(b). 

The different spectral behaviour for ED and MD stems from the physical nature of these multipoles. The MD response is known to be produced by a current loop induced in this particular case in the \textit{xz}-plane. With the decrease in height, the effective space volume pertaining to the current loop decreases, reducing the resonance energy of the MD mode. On the contrary, the ED energy is related to the bulk polarization distributed mostly in the \textit{yx}-plane inside the structure. Consequently, because of the asymmetry of the structure with respect to the \textit{z}-axis, these two multipoles are affected differently as we vary radius and height \cite{Groep_2013}.
   
In order to get a complete picture, we also present the spectra attributed to the simultaneous variation of height and radius in Fig.~\ref{Si_R_and_H_vary}(c). In this case, the blueshift rate is almost the same for the ED and MD. Also, we observe a simultaneous magnitude reduction for both resonances as we decrease the cylinder volume.

Figure SI 1 shows the position of the peaks obtained from Fig.~\ref{Si_R_and_H_vary} for the ED and MD resonances as a function of the geometrical parameters. Here, the circles stand for the resonance peaks at lower energies while the diamonds stand for the resonance peaks at higher energies. It should also be noted that in this plot we have no data pertaining to $R=47$ nm and $H=44$ nm since there is no resonance peak for the ED or the MD in the considered energy range, Fig.~\ref{Si_R_and_H_vary}(c). At 1.07 eV we observe a dip in the dipolar trace for $R=235$ nm $H=220$ nm, which is associated with the excitation of the anapole state \cite{Monticone_2019}. This rather unusual mode stems from the destructive interference between the ED and the toroidal dipole (TD)\cite{Miroshnichenko_2015,Savinov_2019,Krasnok_2019}. This is confirmed by plotting the SCS of the ED and TD, which indeed intersect at this wavelength with a phase difference of $\pi/2$ (data not shown). Interestingly, the anapole state also blueshifts more with decreasing radii, as compared to the decrease in height, since the anapole state appears due to the destructive interference of ED and TD; since the ED shifts more with change in radius, so does the anapole state.
 
\subsection{Influence of the metal disk}

\begin{figure}[h!]
\centering
  \includegraphics[scale=0.85]{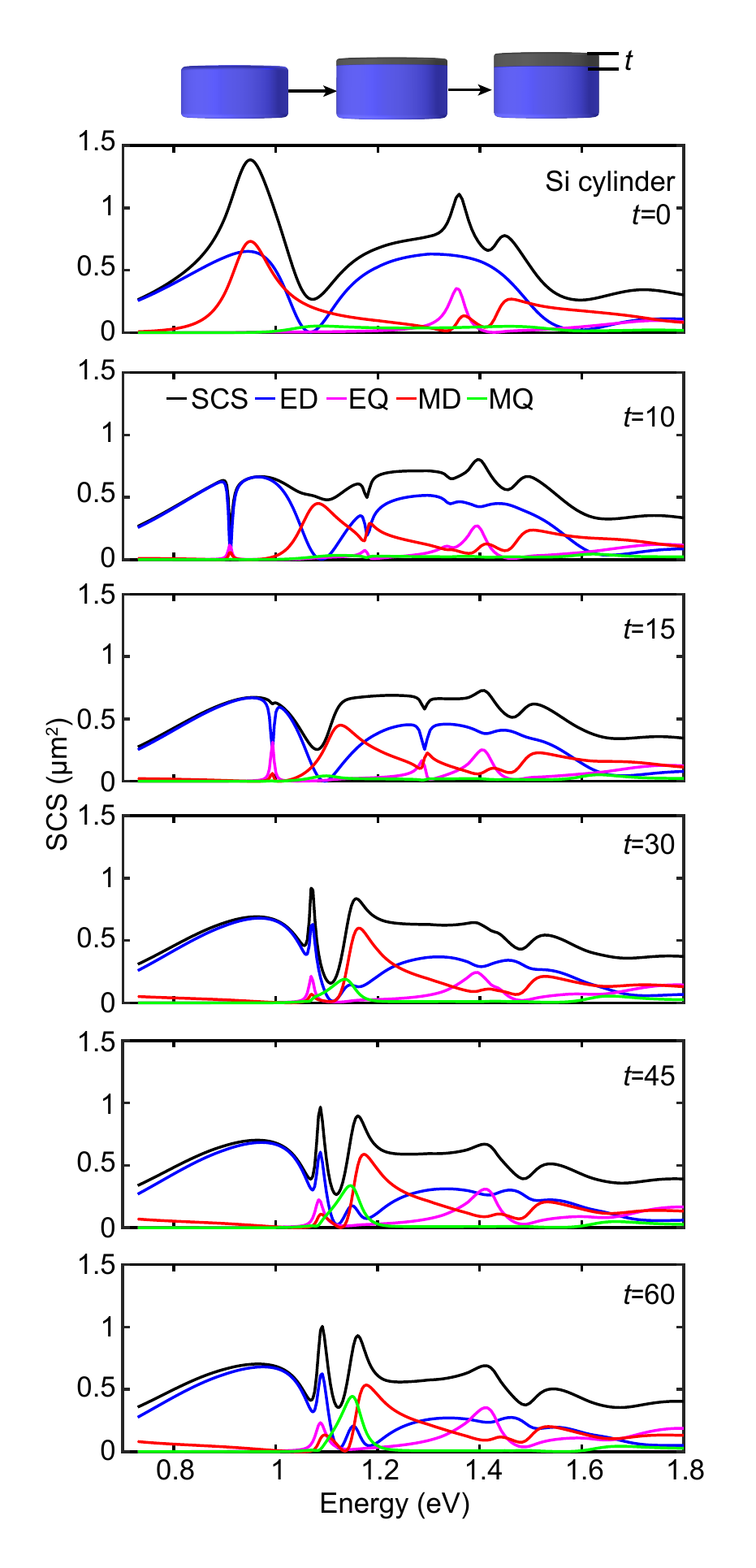}
  \caption{SCS and its Cartesian multipoles decomposition for a Si cylinder ($R=235$ nm, $H=220$ nm) covered with an Ag disk of varying thicknesses $t$. Note that for $t=15$ nm, the SCS at 1.11 eV is solely defined by the MD. All dimensions are in nm.}
\label{SiR235H220_Sp0_AgR235H_vary}
\end{figure}

Having studied the effects caused by increasing the dimensions of the Si cylinder, we now study the effects caused by adding a metallic component in the system, i.e.\ by appending a few nm thick metallic disk on top of the Si cylinder. Indeed, quite significant level of control over the position of the resonances can be achieved by introducing a metal in the system \cite{Li_2020d,Evlyukhin_2015}. We choose a Si cylinder with $H=220$ nm and $R=235$ nm since it has prominent additional multipoles, like MD and EQ. The advantage of involving these multipoles in the analysis can be appreciated in the following where "magnetic light" is achieved. We start with the Ag thickness $t=10$ nm, as this is experimentally achievable \cite{Thyagarajan_2016,Wack_2019,Makela_2017}, and increase the thickness of Ag up to 60 nm \cite{Debdatta_2020}, as shown in Fig.~\ref{SiR235H220_Sp0_AgR235H_vary}.

An addition of metal, albeit extremely thin, drastically modifies the spectrum. In contrast to the dielectric cylinder, where we were able to only manipulate the position and the magnitude of the resonances, we are now able to modify the profile of the resonance itself. Indeed, by inspecting the dipolar trace of the spectrum achieved by adding only 10 nm of Ag we notice sudden dips at 0.91, 1.18 and 1.34 eV, compare $t=0$ and $t=10$ in Fig.~\ref{SiR235H220_Sp0_AgR235H_vary}(b). The dip at 0.91 eV is fairly sharp with a full width at half-maximum of 0.01 eV. The emergence of these dips can be explained by the interference between two ${\pi}$-shifted Cartesian electric dipoles excited in the metal and dielectric parts, giving rise to the effective excitation of the EQ and MD resonances in the far-field \cite{Debdatta_2020}. Quite interestingly, this effect of efficient translation of two ED Cartesian multipoles into EQ and MD is seen only at particular frequencies. Moreover, it becomes less noticeable as the metal thickness increases; possibly because the intrinsic losses in the metal take over.

From the findings for a dielectric cylinder, it is expected to observe a redshift of the MD resonance as the height of the structure is increased. Surprisingly, here the overall profile of the MD spectrum retains its position and magnitude, once the Ag thickness is increased above 15 nm. In the meanwhile the position and the profile of the ED resonance is subject to more noticeable changes. Thanks to these features, almost pure MD resonance can be achieved in this geometry at 1.11 eV, Fig.~\ref{SiR235H220_Sp0_AgR235H_vary} for $t=15$ nm. Similar results have been reported for the geometries of two co-axial metal-dielectric cylinders \cite{Feng_2017} and two metallic cylinders placed one on top of another and separated by a dielectric \cite{Zhang_2019e}. The decoupling of the ED and MD in the range of 1-1.2 eV is also very beneficial for the generation of an excellent anapole state. For example, for Ag thicknesses between $t=30$ nm and $60$ nm in Fig.~\ref{SiR235H220_Sp0_AgR235H_vary} we observe around 1.1 eV negligible ED and MD components, but still a sufficiently strong MQ contribution. It has been shown that, by using radially polarized excitation instead of linearly polarized one, the magnetic contributions can be suppressed \cite{Li_2020d,Das_2015}. This will result in an excellent anapole state making the structure non-radiating at this wavelength. The broad dip in the ED at 1.1 eV for Ag disk of height $t=30$ nm (Fig.~\ref{SiR235H220_Sp0_AgR235H_vary}) is attributed to the anapole state, which leads to the minimum in the ED response and so only a contribution from the MQ is seen \cite{Miroshnichenko_2015}. This anapole state suffers negligible blueshifting with an increase in the thickness of Ag. Also, we noticed that increasing the height of silver above 60 nm barely affects the spectral features of the system. Thus, we take the Si cylinder with 60 nm thick Ag disk as a final structure on which we carry out further modifications to understand the interaction between the Si cylinder and the Ag disk. 
 
\begin{figure}[h!]
\centering
  \includegraphics[width=\textwidth]{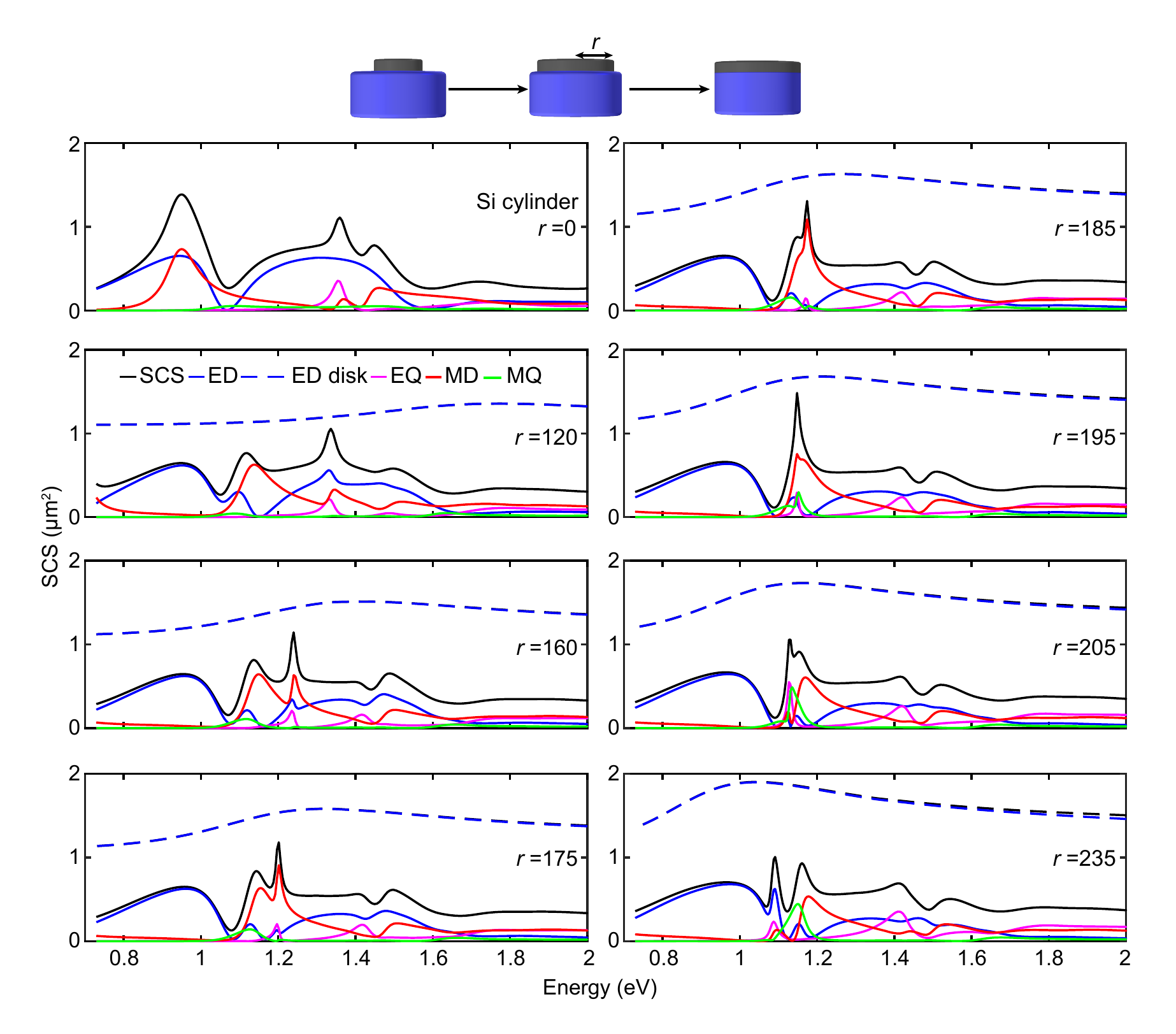}
  \caption{SCS and its Cartesian multipoles decomposition for a Si cylinder ($R=235$ nm, $H=220$ nm) covered with a $t=60$ nm thick Ag disk of varying radii $r$. All dimensions are in nm.}
	\label{SiR235H220_Sp0_AgR_varyH60}
\end{figure}

The spectra reported in Fig.~\ref{SiR235H220_Sp0_AgR235H_vary} have multiple prominent resonant features, which distinguish these hybrid nanostructures from their pure metal or dielectric counterparts. However, sometimes there is a need for a spectrum with only one or two sharp features in the SCS, e.g.\ for sensing when the change in the refractive index of the background analyte is measured by tracking one or two spectral features \cite{Unger_2009a}. In this case, the sharper the resonance, the higher is their figure of merit. This can be achieved in the nanostructure at hand by varying the radius of the Ag disk as shown in Fig.~\ref{SiR235H220_Sp0_AgR_varyH60} for Ag disks with radii between $r=120$ nm and 235 nm (for clarity, the baseline for the Ag disk SCS is shifted by 1${\mu}$m\textsuperscript{2}). We notice that the MD position can be adjusted within the range of 1.08-1.35 eV by changing the Ag disk radius. Specifically, the magnitude of both MD resonances within this region is much stronger than in the remaining part of the spectrum. For $r=160$ nm, we find the MDs at 1.15 and 1.24 eV, Fig.~\ref{SiR235H220_Sp0_AgR_varyH60}. For larger Ag radii, the lower energy MD remains around 1.15 eV, while the higher energy MD slightly redshifts for $r=175$ nm, and then begins to fuse with the low energy MD, producing a very strong spectral feature for $r=185$ nm. This feature becomes even more prominent for $r=195$ nm, where it dominates the entire SCS, Fig.~\ref{SiR235H220_Sp0_AgR_varyH60}. This peak can also be explained by the interference between two ${\pi}$-shifted Cartesian electric dipoles excited in the metal and dielectric parts giving rise to effective excitation of EQ and MD resonances in the far field. The MD resonance arising from this phenomenon interferes with the MD from the Si cylinder and reduces the net MD of the system \cite{Debdatta_2020}. Such a very sharp feature could be very useful for sensing applications as it will give high figure of merit \cite{Unger_2009a, Spackova_2016}. Beyond $r=195$ nm, this sharp feature at 1.15 eV starts again to split into two peaks as shown in Fig.~\ref{SiR235H220_Sp0_AgR_varyH60} for $r=205$ and 235 nm. 
 
\subsection{Coupling between the dielectric cylinder and the metal disk}
\begin{figure}[h!]
\centering
  \includegraphics[width=\textwidth]{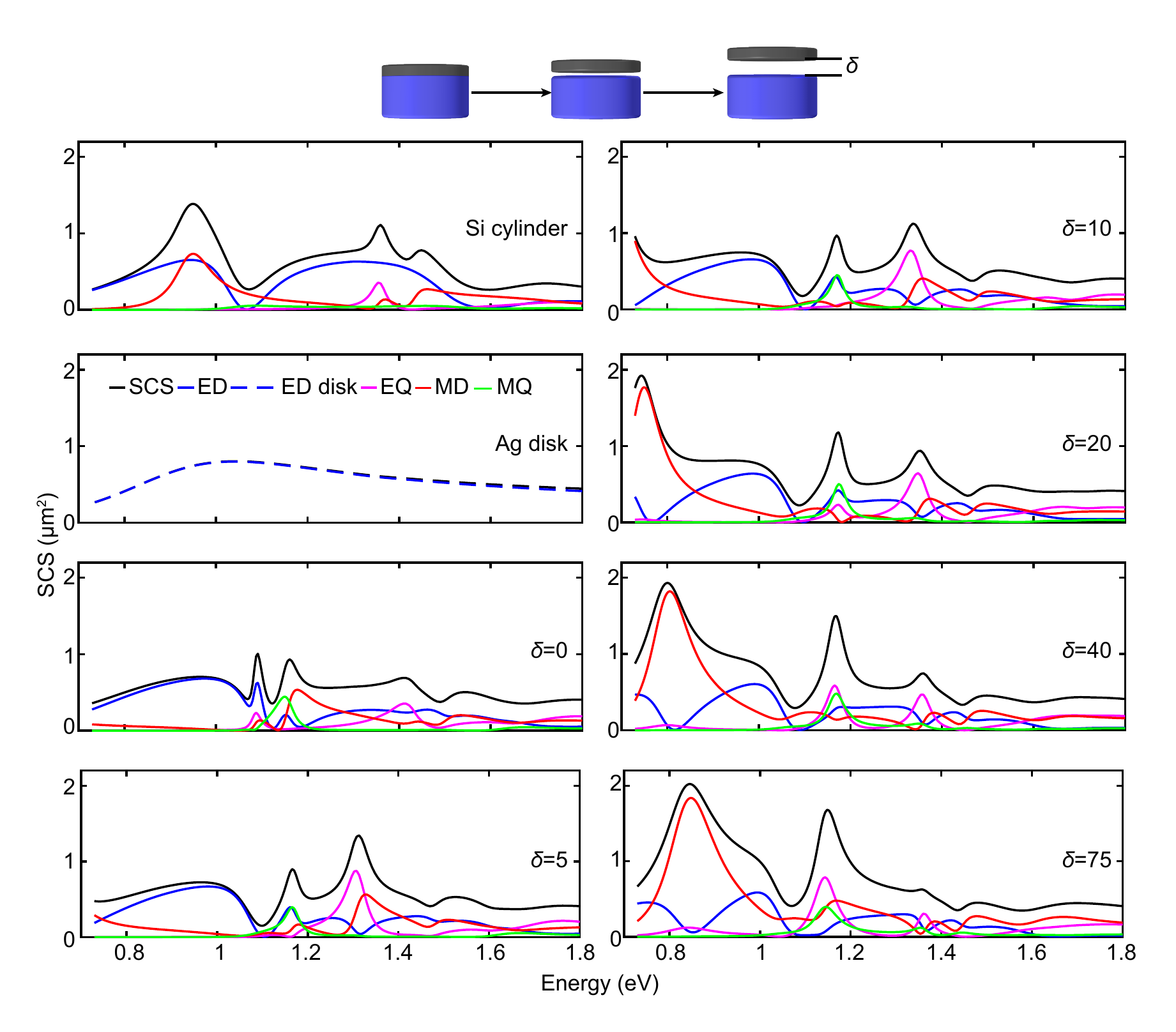}
  \caption{SCS and its Cartesian multipoles decomposition for a Si cylinder ($R=235$ nm, $H=220$ nm) covered with an Ag disk ($t=60$ nm, $r=235$ nm) placed at a spacing distance $\delta$. The first two panels show the response for the individual cylinder and disk.}
\label{SiR235H220_Sp_vary_AgR235H}
\end{figure}

Finally, we investigate an additional control over the spectrum offered by this hybrid nanostructure by introducing a spacing of varying thickness $\delta$ between the Si cylinder and the Ag disk. For the ease of understanding, we consider an air spacer; a dielectric spacer will merely increase the coupling as is the case for a dielectric loaded plasmonic structure \cite{Abasahl_2014}. This study complements the one done previously for Si cylinder sandwiched between two Ag disks \cite{Zhang_2019e}. In Fig.~\ref{SiR235H220_Sp_vary_AgR235H} we slowly increase the spacing from $\delta=5$ nm, which is achievable with current standards of nanofabrication \cite{Ma_2019a}, to $\delta=75$ nm. 

The most prominent feature that can be noticed as we increase the spacing is the appearance of a broader resonance in the SCS at 0.74 eV due to the emergence in the spectral window of a very strong MD resonance for a separation larger than $\delta=10$ nm. Such a feature in the near infrared can be useful for applications in surface enhanced infrared absorption \cite{Adato_2013a}. It can also be used in applications requiring only the magnetic response from the structure. However, overall, the spectral features observed for large spacings are relatively broad and will not be well suited for practical applications. Another feature in the spectrum attracted our attention and is worth mentioning. The two peaks at 1.09 eV and 1.16 eV in the SCS shown in Fig.\ \ref{SiR235H220_Sp_vary_AgR235H} for $\delta=0$, are relatively sharp. By increasing the spacing between dielectric and metal, the peak at higher energy becomes less pronounced while the magnitude of that at lower energy significantly increases. This is in contrast to the effect shown in Fig.\ \ref{SiR235H220_Sp0_AgR235H_vary}, where adding Ag on top of the Si cylinder introduced sharp small dips in the ED spectra in the range of 1-1.5 eV which affected the SCS in this range likewise. 

\begin{table}
\begin{center}

\caption{The spectral changes due to changes in the Si cylinder radius $R$ and height $H$, the metal disk thickness $t$ and radius $r$, or the spacing distance $\delta$. The symbols indicate: $\leftarrow$ = redshift, $\rightarrow$ = blueshift, $\uparrow$ = increase, $\downarrow$ = decrease, $\times$ = no significant change.}

\label{Table_summarise}
\begin{tabular}{|c|c|c|c|c|c|} 
\hline
\multirow{2}{*}{Nanostructure} & \multirow{2}{*}{Parameter} & \multicolumn{2}{c| } {ED} & \multicolumn{2}{c|} {MD} \\
\cline{3-6}
&   & Energy & Amplitude & Energy & Amplitude \\
\hline
\multirow{3}{*}{Si cylinder} & $R$ $\uparrow$ & $\leftarrow$ & $\uparrow$ & $\leftarrow$ &  $\times$\\ 
\cline{2-6}
& $H$ $\uparrow$ & $\times$ & $\uparrow$ & $\leftarrow$ & $\uparrow$\\ 
\cline{2-6}
& $R$ \& $H$ $\uparrow$ & $\leftarrow$ & $\uparrow$ & $\leftarrow$ & $\uparrow$\\
\hline
\multirow{3}{*}{Hybrid structure} & $t$ $\uparrow$ & $\rightarrow$ & $\times^{1}$ & $\times$ & $\times$\\ 
\cline{2-6}
& $r$ ${\uparrow}$ & $\times$ & $\times$ & $\times$ & $\times^2$\\ 
\cline{2-6}
& $\delta$ $\uparrow$ & $\times$ & $\times^3$ & $\rightarrow$ & $\times$ \\ 
\hline
\end{tabular}
\end{center}
$\,$\\\footnotesize{$^{1}$ Sharp optical features.}\\
\footnotesize{$^{2}$ Both MD fuse to give rise to a sharp peak at $r= 195$ nm and then split again.}\\
\footnotesize{$^{3}$ Strong interaction between ED and MD.}
\end{table}

\section{Conclusion}

To conclude, a hybrid dielectric--metal nanostructure provides many degrees of freedom to control different spectral features, including pure magnetic modes and anapoles, as well as extremely narrow resonances. The results of the geometrical variations are summarised in Table.~\ref{Table_summarise}. We see that, depending on the geometry, the ED and MD resonances can be spectrally overlapped or separated and the knowledge gathered in this study could be combined with advanced design approaches to develop systems that support a given set of multipolar characteristics\cite{Blanchard-Dionne_2020, Qin_2019a}. Overall, hybrid metal-dielectric nanostructures provide both a very high degree of freedom and a good control for engineering resonant systems with tailored spectral characteristics.

\begin{backmatter}
\bmsection{Funding}
Funding from the Swiss National Science Foundation (project 200021-162453) and the European Research Council (ERC-2015-AdG-695206 Nanofactory) is gratefully acknowledged.

\bmsection{Acknowledgments}
It is a pleasure to acknowledge stimulating discussions with G.\ D.\ Bernasconi during the initial phase of this project.

\bmsection{Disclosures}
The authors declare no conflicts of interest.

\bmsection{Data availability} Datasets underlying the results presented in this paper are available in Refs. \cite{Kiselev_VSH, Ag_data_Debdatta}.

\bmsection{Supplemental document}
See Supplement 1 for supporting content. 

\end{backmatter}
\bibliography{reference}

\clearpage
\section{Supplementary Information: Variation of peaks of ED and MD resonances from Fig. 1}
\renewcommand{\figurename}{Figure}
\begin{figure}[hp!]
\centering
  \includegraphics[scale=0.9]{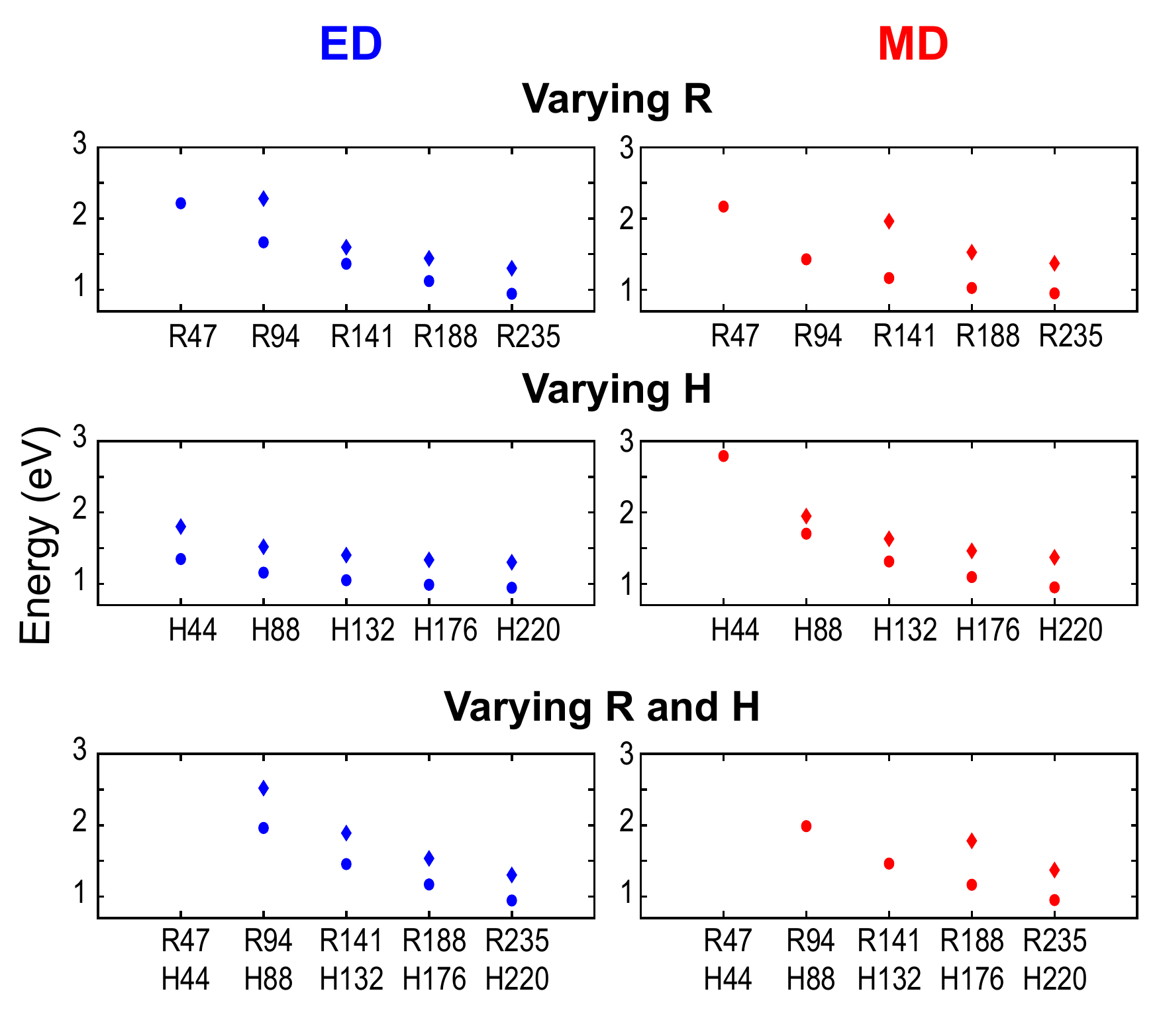}
  \caption{ Resonance peak positions of the ED and MD varying with the change in dimensions. The circles refer to peaks at lower energy while the diamonds refer to peaks at higher energy. 
}
 \label{Variation_of_Si_summary}
\end{figure}

\end{document}